\documentclass[aps,prd,groupedaddress,preprint,eqsecnum,nofootinbib]{revtex4}
\usepackage{amsmath,amssymb}
\usepackage{float,cancel}
\usepackage{mathtools}
\usepackage{xcolor}
\usepackage{graphicx}
\usepackage{longfigure}
\usepackage{capt-of} 
\usepackage{pdfpages}
\usepackage{bm}
\usepackage{subcaption}
\usepackage{hyperref}
\usepackage{hyperref}
\usepackage{tikz}
\usepackage{tikz-feynman}
\usepackage{empheq}
\newcommand{\f}{\frac}

\newcommand{\be}{\begin{equation}}
\newcommand{\ee}{\end{equation}}
\newcommand{\ben}{\begin{eqnarray}\displaystyle}
\newcommand{\een}{\end{eqnarray}}
\newcommand{\non}{\nonumber}

\def\be{\begin{equation}}
\def\ee{\end{equation}}

\def\b{\beta}

\def\f{\phi}

\def\bg{\bar{g}}
\def\beq{\begin{eqnarray}}\def\eeq{\end{eqnarray}}
\def\ba#1\ea{\begin{align}#1\end{align}}
\def\bg#1\eg{\begin{gather}#1\end{gather}}
\def\bm#1\em{\begin{multline}#1\end{multline}}
\def\bmd#1\emd{\begin{multlined}#1\end{multlined}}
\def\f{\frac}

\def\b{\beta}

\def\D{\Delta}

\def\p{\phi}

\def\({\left(}
\def\){\right)}
\def\[{\left[}
\def\]{\right]}

\def\b{\beta}

\def\p{\partial}

\def\f{\frac}

\def\D{\Delta}

\hypersetup{
	colorlinks=true,
	linkcolor=blue,
	urlcolor=magenta,
	citecolor=red,
}

\allowdisplaybreaks
\def\be{\begin{equation}}
	\def\ee{\end{equation}}

\begin{document}
\hfuzz 9pt
\title{All loop soft photon theorems and higher spin currents on the celestial sphere}
\author{Shamik Banerjee$^{1,2}$}
\author{Raju Mandal$^{1,2,3}$}
\author{Biswajit Sahoo$^4$}\email{banerjeeshamik.phy@gmail.com, rajuphys002@gmail.com, biswajit.sahoo@kcl.ac.uk}
\affiliation{$^1$National Institute of Science Education and Research (NISER), Bhubaneswar 752050, Odisha, India}
\affiliation{$^2$Homi Bhabha National Institute, Training School Complex, Anushakti Nagar, Mumbai 400094, India}
\affiliation{$^3$Indian Institute of Technology Ropar, Punjab 140001, India}
\affiliation{$^4$Department of Mathematics, King’s College London, Strand, London WC2R 2LS, United Kingdom}
\begin{abstract} 
Soft factorization theorems can be reinterpreted as Ward identities for (asymptotic) symmetries of scattering amplitudes in asymptotically flat space-time. In this paper we study the symmetries implied by the all loop soft photon theorems when all the charged particles are highly energetic and the relation $\omega << m << E$ holds where $E$ is the typical energy of a charged particle, $m$ is the typical mass and $\omega$ is the soft photon energy. Loop level soft theorems are qualitatively different from the tree level soft theorems because loop level soft factors contain multi-particle sums. If we want to interpret them as Ward identities or define celestial OPE between between soft and hard operators then we need to introduce additional fields which live on the celestial sphere but do not appear as asymptotic states in any scattering experiment. For example, if we want to interpret the one-loop exact $\mathcal{O}(\ln\omega)$ soft theorem for a positive helicity soft photon (with energy $\omega$) as a Ward identity then we need to introduce a pair of antiholomorphic currents on the celestial sphere which transform as a doublet under the $SL(2,\mathbb{R})_{R}$. We call them dipole currents because the corresponding charges measure the monopole and the dipole moment of an electrically charged particle on the celestial sphere. More generally, the soft photon theorem at $\mathcal{O}(\omega^{2j-1}(\ln\omega)^{2j})$ for every $j\in \frac{1}{2}\mathbb{Z}_+$ gives rise to $(2j+1)$ antiholomorphic currents which transform in the spin-$j$ representation of the $SL(2,\mathbb{R})_{R}$. These currents exist in the quantum theory because they follow from loop level soft theorems. We argue that under certain circumstances the (classical) algebra of the higher spin currents is the wedge subalgebra of the $w_{1+\infty}$.

\end{abstract}


\maketitle
\tableofcontents

\section{Introduction}
Soft factorization theorems can be reinterpreted as Ward identities for symmetries of scattering amplitudes in asymptotically flat space-time. This plays an important role in Celestial Holography \cite{Strominger:2017zoo,Pasterski:2016qvg,Banerjee:2018gce} which posits that the holographic dual of quantum theory of gravity in asymptotically flat four dimensional space time is a CFT$_2$ living on the celestial two-sphere. The symmetries of the scattering amplitudes arising out of soft theorems are global symmetries of the Celestial CFT$_2$ and in many cases the global symmetry is infinite dimensional \cite{Strominger:2013lka,Barnich:2009se,Banerjee:2020zlg,Guevara:2021abz,Strominger:2021mtt,Himwich:2021dau,Adamo:2021lrv,Freidel:2021ytz,Stieberger:2023fju,Melton:2024akx}. This is a powerful statement because one can constrain four dimensional scattering amplitudes \cite{Banerjee:2020zlg,Banerjee:2023zip,Fan:2022vbz,Costello:2023vyy} using the infinite dimensional global symmetries. Moreover, understanding how these symmetries are realized in celestial CFT$_2$ could shed light on the connection between celestial holography and string theory in asymptotically flat spacetime \cite{Kervyn:2025wsb,Donnay:2023kvm,Banerjee:2025tec}. 

In this paper, we study the symmetry interpretation of the soft photon theorems that remain valid even at loop level. Consequently, the symmetries implied by these soft theorems are suppose to be symmetries of the holographic dual celestial CFT$_2$, persisting even after quantum corrections are taken into account. It is well known that the leading soft photon theorem, derived in 
\cite{Gell-Mann:1954wra,Low:1954kd,Weinberg:1965nx}, is universal and exact at 
tree level, receiving no loop corrections\footnote{Up to a subtlety recently 
pointed out in \cite{Ma:2023gir}, arising from a special class of three-loop 
diagrams in the presence of massless charged particles.}. However, since four-dimensional $S$-matrices are infrared divergent due to the long-range interactions mediated by massless particles, deriving finite subleading corrections has long remained a challenge \cite{Cachazo:2014dia,Bern:2014oka,He:2014bga}. By interpreting soft factors as ratios of $S$-matrices with and without an external soft photon, and by systematizing the exponentiation of infrared divergences, it is possible to extract well-defined subleading soft theorems from the IR-finite parts of the ratios of $S$-matrices. This program was successfully carried out in \cite{Sahoo:2018lxl}, with a proof of universality subsequently established in \cite{Krishna:2023fxg}. The resulting subleading soft photon theorem appears at $\mathcal{O}(\ln \omega)$ in the soft photon energy $\omega$ and is one loop exact. The $\mathcal{O}(\ln\omega)$ soft factor consists of two distinct contributions, termed the ``classical'' and ``quantum'' parts in \cite{Sahoo:2018lxl}. The classical contribution encodes information about the classical electromagnetic waveform in the low-frequency expansion, as demonstrated through independent classical analyses in \cite{Laddha:2018myi,Saha:2019tub}. By contrast, a clear physical interpretation of the quantum contribution remains elusive to date. In this paper we aim to understand it as a Ward identity in the framework of celestial CFT and to uncover the associated symmetry structure. 


From a four-dimensional bulk perspective, the \textit{classical} part of the subleading soft photon theorem at $\mathcal{O}(\ln\omega)$ has already been understood as a Ward identity associated with a large gauge symmetry.\footnote{The Celestial OPE of gluons and gravitons at one-loop were studied in \cite{Bhardwaj:2022anh,Krishna:2023ukw}.} This symmetry acts on charged matter fields as a local enhancement of global phase rotations, referred to as ``superphaserotations'' in \cite{Campiglia:2019wxe,AtulBhatkar:2019vcb,Choi:2025mzg}. In the present work we provide an interpretation of the \textit{quantum} contribution to the $\mathcal{O}(\ln\omega)$ soft photon theorem in QED as a symmetry of the dual Celestial CFT$_2$ in the \textit{high energy limit}. Furthermore, we analyze higher subleading corrections to the soft photon theorem at orders $\omega^n(\ln\omega)^{n+1}$ as conjectured and explicitly derived up to sub-subleading order in \cite{Sahoo:2020ryf,Karan:2025ndk}, and argued to be $(n+1)$-loop exact. In the high energy limit the classical part of the soft theorem is subleading to the quantum part. 


Loop level soft theorems are qualitatively different from the tree level ones. It turns out that if we want to interpret them as Ward identities of symmetries in a Celestial CFT$_2$ or define a Celestial OPE between a hard and soft operators then we need to introduce additional fields on the celestial sphere which do \textit{not} appear as asymptotic states in scattering experiments. In particular, they are \textit{not} related to the soft gauge bosons in any obvious manner. Now since there are an infinite number of loop level soft photon theorems we get an infinite number of chiral currents in this way. For every $j\in\frac{1}{2}\mathbb{Z_+}$ the soft theorem at $\mathcal{O}(\omega^{2j-1}(\ln\omega)^{2j})$ gives rise to $(2j+1)$ chiral currents which transform in the finite dimensional spin-$j$ representation of the $SL(2,\mathbb{R})_{R}$. The charges corresponding to the $j=1/2$ currents measure the monopole and the dipole moment of an electrically charged particle on the celestial sphere and hence called a dipole current. The other spin-$j$ currents can be constructed by taking the (normal ordered) product of $2j$ dipole currents. Depending on the value of a real parameter $k$ which appears in the OPE of the dipole currents, the algebra of the higher spin currents can be Abelian ($k=0$) or the wedge subalgebra ($k<<1$) of the $w_{1+\infty}$.  

Appearance of higher spin currents is not new in the context of Celestial holography. Both soft gluons and gravitons generate infinite dimensional higher spin algebras known as the $S$-algebra and the wedge subalgebra of $w_{1+\infty}$ \cite{Strominger:2021mtt,Adamo:2021lrv,Freidel:2021ytz,Himwich:2021dau}. However, the infinite number of higher spin currents we get from the loop level soft photon theorems and the symmetries they generate are robust against quantum corrections, at least in the high energy limit. It remains to be seen how one can exploit this feature to constrain scattering amplitudes or relate celestial holography to string theory in asymptotically flat space-time.\footnote{We have more to say on this in an upcoming paper.} \\



\textbf{Note Added:}
In a previous version of the paper instead of the \textit{high energy} limit we talked about the \textit{massless} limit of the QED. Although in many cases these two limits are synonymous, in the present context this is not so. There are two reasons. Firstly, there are two low energy scales in the problem. One is the mass $m$ of the charged particle and the other one is the energy $\omega$ of the soft photon. The results of \cite{Sahoo:2018lxl} were derived under the assumption that $\omega << m$ and so massless limit cannot be taken. In other words, $\omega$ is the smallest energy scale in the problem and this should be strictly maintained at every stage of any limiting procedure. The second, and more fundamental, reason is that in the massless QED the physical charge goes to zero in the infrared and therefore soft symmetries cannot be defined. We are grateful to Andrew Strominger for pointing this out to us. 

The high energy limit that is discussed in the following section does not suffer from any of these problems and all the technical results presented in the previous version remain unchanged.

\section{All loop soft theorems and high energy limit} Consider the scattering of massive charged particles in a generic theory of quantum electrodynamics in four spacetime dimensions, where, apart from electromagnetism, no other long-range interaction is present. The soft photon theorem valid at all order in perturbation theory relating the scattering amplitude $\mathcal{A}^{(\text{em})}_{N+1}$, involving $N$ number of finite-energy charged particles with charges $\lbrace e_a \rbrace$ and momenta $\lbrace p_a \rbrace$, and one outgoing soft photon with polarization $\varepsilon^h_\mu$, helicity $h$, and momentum $k^\mu = -\omega \mathbf{n}^\mu$ (where $\omega$ is the photon energy and $\mathbf{n}^\mu$ is the unit null vector specifying the emission direction), to the scattering amplitude $\mathcal{A}^{(\text{em})}_{N}$ without the external photon reads:\footnote{As is well known, in four spacetime dimensions both $\mathcal{A}^{(\text{em})}_{N+1}$ and $\mathcal{A}^{(\text{em})}_{N}$ are infrared (IR) divergent. However, since the soft factor appears as a ratio of these amplitudes, it is IR finite in the sense that the soft theorem in~\eqref{eq:Soft_expansion} relates the IR-finite hard parts of the two amplitudes, following the arguments of~\cite{Sahoo:2018lxl,Arkani-Hamed:2020gyp,Krishna:2023fxg}.}${}^{,}$\footnote{In this paper, we adopt the mostly positive convention for the Minkowski metric. The charges and energies of incoming particles are taken with their physical signs, while those of outgoing particles acquire an additional negative sign, ensuring that charge and momentum conservation are expressed as $\sum\limits_{a} e_a = 0$ and $\sum\limits_{a} p_a = 0$, respectively.}
\ben
\mathcal{A}^{\text{(em)}}_{N+1}\overset{\omega\rightarrow 0}{=}S_{\text{em}}\left(\lbrace e_a,p_a\rbrace ,(\varepsilon^h,k)\right)\,  \mathcal{A}^{\text{(em)}}_N\, +\, \mathcal{R}^{\text{(em)}}_{N+1}, \label{eq:Soft_expansion}
\een
where $S_{\text{em}}$ represents the universal soft photon factor determined only in terms of the scattering data specified in its argument and $ \mathcal{R}^{\text{(em)}}_{N+1}$ represents the non-factorized part of $\mathcal{A}^{\text{(em)}}_{N+1}$ in the soft limit.\footnote{For the all-loop resumed S-matrix in a generic theory of QED, $\mathcal{R}^{\text{(em)}}_{N+1}$ cannot be related to $\mathcal{A}^{\text{(em)}}_{N}$, and hence it does not factorize. However, if we restrict to the tree-level S-matrix, then $\mathbb{R}_{\text{em},N+1}^{(0,0)}$ factorizes and yields the well-known Low’s soft photon theorem, with a specific contribution from the non-minimal coupling through the field strength \cite{Low:1954kd,Elvang:2016qvq}. At one-loop order, $\mathbb{R}_{\text{em},N+1}^{(0,0)}$ acquires non-universal contributions originating from different regions of the loop diagrams which fails to factorize \cite{Krishna:2023fxg}.} They have the following expansion in the soft limit:
\be \label{eq:S_em_expansion}
\begin{gathered}
S_{\text{em}}\left(\lbrace e_a,p_a\rbrace ,(\varepsilon^h,k)\right)\overset{\omega\rightarrow 0}{=}\sum_{n=-1}^{\infty} \mathbb{S}_{\text{em}}^{(n)} \times \omega^n(\ln\omega)^{n+1},\\
\mathcal{R}^{(\text{em})}_{N+1} \overset{\omega\rightarrow 0}{=} \sum_{n=0}^\infty \sum_{t=0}^n \mathbb{R}_{\text{em},N+1}^{(n,t)} \, \omega^n(\ln\omega)^{t}.
\end{gathered}
\ee
The soft factor $\mathbb{S}_{\text{em}}^{(n)}$ has been argued to be $(n+1)$-loop exact. 
The soft factor $\mathbb{S}_{\text{em}}^{(-1)}$ was derived in 
\cite{Gell-Mann:1954wra,Low:1954kd,Weinberg:1965nx}, 
$\mathbb{S}_{\text{em}}^{(0)}$ in \cite{Sahoo:2018lxl}, and 
$\mathbb{S}_{\text{em}}^{(1)}$ in \cite{Sahoo:2020ryf}, for the scattering of 
massive charged particles. The universality of 
$\mathbb{S}_{\text{em}}^{(0)}$ was established in \cite{Krishna:2023fxg}.

\subsection{High energy limit}

In this paper we want to study the soft expansion \eqref{eq:S_em_expansion} in the high energy limit 
\be\label{domain}
\boxed{
\omega^2 << m^2 << |p_a . p_b|\sim E^2},\quad  \forall  a\ne b,
\ee
where $m$ represents the typical mass of a charged particle, $E$ is the characteristic energy scale of the hard charged particles and $\omega$ is the energy of the soft photon. 


In the high energy limit \eqref{domain} the expressions for the soft factors simplify considerably. For example, the $\mathcal{O}(\ln\omega)$ soft factor given by \cite{Sahoo:2018lxl}
\be
\begin{gathered}
\mathbb{S}^{(0)}_{\text{em}} =  \frac{i}{4\pi} \sum_{a=1}^N \sum^N_{\substack{b=1\\b\ne a, \eta_a\eta_b=1}} e_a^2 e_b \frac{\epsilon_\mu \mathbf{n}_\rho}{p_a.\mathbf{n}} \frac{m_a^2 m_b^2 \[ p^\mu_a p^\rho_b - p^\mu_b p^\rho_a\]}{\[ (p_a.p_b)^2 - m_a^2 m_b^2\]^{\frac{3}{2}}} \\
+\frac{1}{8\pi^2} \sum^N_{\substack{a,b=1\\ a\ne b}} e_a^2 e_b \log\[\frac{p_a.p_b + \sqrt{(p_a.p_b)^2- m_a^2 m_b^2}}{p_a.p_b - \sqrt{(p_a.p_b)^2- m_a^2 m_b^2}}\] \frac{m_a^2m_b^2}{\((p_a.p_b)^2 - m_a^2 m_b^2\)^{3/2}}\( -\epsilon.p_b + \frac{\epsilon.p_a}{\mathbf{n}.p_a} \mathbf{n}.p_b\) \\
- \frac{1}{4\pi^2}\sum^N_{\substack{a,b=1\\ a\ne b}} e_a^2 e_b \frac{p_a.p_b}{(p_a.p_b)^2 - m_a^2 m_b^2} \( -\epsilon.p_b + \frac{\epsilon.p_a}{\mathbf{n}.p_a} \mathbf{n}.p_b\)
\end{gathered}
\ee
can be expanded in the limit \eqref{domain} and the result is given by 
\be\label{rational}
\begin{gathered}
\mathbb{S}^{(0)}_{\text{em}}= - \frac{1}{4\pi^2}\sum_{\substack{a,b=1\\ b\neq a}}^N  \frac{e^2_{a}e_{b}}{p_{a}.p_{b}}\frac{\varepsilon_\mu \mathbf{n}_\nu}{p_{a}.\mathbf{n}}\left(p_{a}^\mu p_{b}^\nu -p_{a}^\nu p_{b}^\mu\right) + \mathcal{O}\( \frac{m^4}{E^5} \log\frac{m^4}{E^4}, \frac{m^4}{E^5}\).
\end{gathered}
\ee
We observe that the soft factor $\mathbb{S}^{(0)}_{\text{em}}$ contains a leading contribution of order $\mathcal{O}(1/E)$ which is a rational function of the hard particle momenta. In addition, it includes power-law and logarithmic correction terms, as indicated inside the parentheses of the $\mathcal{O}(\cdots)$ symbol. These terms are subleading in the limit defined in \eqref{domain}. Throughout this paper, we restrict our attention to the leading rational contribution appearing in \eqref{rational}. 

Now in QED the high energy limit is not really well defined because of the existence of the Landau pole. We assume that the energies of the hard particles are well below that but they are much greater than the typical mass of a charged particle so that we can neglect the logarithmic and power law suppressed terms in \eqref{rational} to a good approximation. However, it should be kept in mind that there is no region of parameter space where the correction terms are zero. It will be very interesting to study the soft symmetry interpretation of the subleading terms in the high energy expansion. We leave it for future investigation. 

Now motivated by the analysis in appendix~\ref{S:appendix} and inspired by refs.~\cite{Sahoo:2020ryf,Karan:2025ndk} we can write down the expressions for the \textit{leading terms} of the logarithmic soft theorems in the high energy limit \eqref{domain} as
\ben\label{eq:soft_photon_coeff}
\begin{gathered}
\mathbb{S}^{(-1)}_{\text{em}}=-\sum_{a=1}^N e_a\frac{\varepsilon^h_\mu p_a^\mu}{p_a.\mathbf{n}},\\
\mathbb{S}^{(n)}_{\text{em}}= - \frac{1}{(n+1)!(4\pi^2)^{n+1}}\sum_{\substack{a,b=1\\ b\neq a}}^N  \frac{e^2_{a}e_{b}}{p_{a}.p_{b}}\frac{\varepsilon^h_\mu \mathbf{n}_\nu}{p_{a}.\mathbf{n}}\left(p_{a}^\mu p_{b}^\nu -p_{a}^\nu p_{b}^\mu\right)\left( \sum_{\substack{c=1\\ c\neq a}}^N \frac{e_{a}e_{c}}{p_{a}.p_{c}}(p_{c}.\mathbf{n})\right)^n,
\end{gathered}
\een
for $n\geq 0$. The $\mathbb{S}^{(n)}_{\text{em}}$ for $n\geq 0$ presented above are intrinsically quantum contributions to the soft photon theorems since the classical contributions are \textit{subleading} in the high energy limit.\footnote{The classical contribution to $\mathbb{S}^{(n)}_{\text{em}}$ for the scattering of massive charged particles was derived in \cite{Karan:2025ndk}. It vanishes in the massless limit, which is physically expected since, even in the presence of a long-range electromagnetic force, massless charged particles always move at the speed of light in flat spacetime, do not accelerate, and therefore do not radiate.} The terminology “classical” and “quantum” follows from \cite{Sahoo:2018lxl}.  From the soft expansion \eqref{eq:Soft_expansion} and \eqref{eq:S_em_expansion}, the soft factor coefficient $\mathbb{S}_{\text{em}}^{(n)}$ can be extracted by using the following projection operator
\be\label{projection}
\mathbb{S}_{\text{em}}^{(n)}=\f{1}{(n+1)!}\lim_{\omega\rightarrow 0}\f{d^{n+1}}{d\omega^{n+1}}\omega\prod_{\ell=0}^{n+1}\f{1}{\ell!}\left(\omega \f{d}{d\omega}-(\ell-1)\right)^\ell\left[\f{\mathcal{A}^{\text{(em)}}_{N+1}}{{\mathcal{A}^{\text{(em)}}_{N}}}\right],
\ee
where the projection operator for $n=0$ was already found in \cite{Campiglia:2019wxe}.

\section{Soft theorems in conformal primary basis}
In order to interpret the soft theorems \eqref{eq:soft_photon_coeff} as Ward identities in a CFT$_2$ living on the celestial sphere we transform to the conformal primary basis. Since we are working in the high energy limit we can treat the hard particles as (effectively) massless because the finite mass correction is suppressed by powers of $m/E$ and hence small according to our assumption. This allows us to parametrize the soft momentum $k^{\mu}$, positive helicity photon polarization vector $\varepsilon^{+\mu}(k)$ and the momentum $p^{\mu}$  of a hard particle as
\ben
\begin{gathered}
k^\mu =\eta\omega\, \mathbf{n}^\mu(w,\bar{w}), \ \omega\rightarrow 0+ \, , \\
\mathbf{n}(w,\bar{w})=\big(1+w\bar{w},w+\bar w,-i(w-\bar w),1-w\bar w\big),\\
\quad \varepsilon^{+\mu}(k) =\f{1}{\sqrt{2}}(\bar w,1,-i,-\bar w),\\
p^\mu=\eta \omega \mathbf{n}(z,\bar z),
\end{gathered}
\een
where $\eta=\pm 1$ for an incoming (outgoing) particle and $(w,\bar{w})((z,\bar{z}))$ represent the (stereographic) coordinates of the Celestial sphere. We also define celestial primary field $\phi^{(\eta)}_{h, \bar h}(z,\bar z,e)$ by Mellin transforming the creation and annihilation operators $b^\dagger$ and $b$ as \cite{Pasterski:2016qvg}
\ben
\begin{gathered}
\phi^{(+1)}_{h,\bar h}(z,\bar{z},e) =  \int_0^\infty d\omega\, \omega^{\Delta-1}\, b^\dagger(p,\sigma,e),\\
\phi^{(-1)}_{h,\bar h}(z,\bar{z},e) =  \int_0^\infty d\omega\, \omega^{\Delta-1}\, b(p,\sigma,e),
\end{gathered}
\een
where $\Delta$ is in general a complex number, $\sigma \in \f{1}{2}\mathbb{Z}$ is the helicity of the particle and $e$ is the electric charge. One can show that under Lorentz transformation which acts on the Celestial sphere as the group ($SL(2,\mathbb{C})$) of global conformal transformations, the field $\phi^{(\eta)}_{h,\bar h}(z,\bar z,e)$ transforms as a conformal primary with holomorphic and anti-holomorphic conformal dimensions given by  $h=\f{1}{2}(\Delta+\sigma)$ and $\bar{h}=\f{1}{2}(\Delta-\sigma)$, respectively. 

Now after transforming to the conformal primary basis, the soft photon theorems in \eqref{eq:soft_photon_coeff} can be expressed as the following identities between celestial correlation functions:
\ben
&&\left\langle \mathcal{S}^1(w,\bar{w}) \prod_{i=1}^N \phi^{(\eta_i)}_{\D_i}(z_i,\bar{z}_i,e_i) \right\rangle = \sum_{a=1}^N  \f{e_a}{w-z_a}\quad \left\langle \prod_{i=1}^N \phi^{(\eta_i)}_{\D_i}(z_i,\bar{z}_i,e_i) \right\rangle , \label{eq:Mellin_leading_soft_photon_thm}\\
&&\left\langle \mathcal{S}^{1-2j}(w,\bar{w}) \prod_{i=1}^N \phi^{(\eta_i)}_{\D_i}(z_i,\bar{z}_i,e_i) \right\rangle
= {\frac{1}{(2j)!}} \sum_{a=1}^N \f{e_a^{2j+1}\eta_a^{2j}}{w-z_a}\Bigg[ \sum_{\substack{b=1\\ b\neq a}}^N  e_b\,  \f{\bar w-\bar z_b}{\bar z_a-\bar z_b}\Bigg]\non\\
&&\times\Bigg[ \sum_{\substack{c=1\\ c\neq a}}^N e_c\, \f{(w-z_c)(\bar w-\bar z_c)}{(z_a-z_c)(\bar z_a-\bar z_c)}\Bigg]^{2j-1} \left\langle \phi^{(\eta_a)}_{\D_a-2j}(z_a,\bar{z}_a,e_a) \prod_{\substack{i=1\\ i\neq a}}^N \phi^{(\eta_i)}_{\D_i}(z_i,\bar{z}_i,e_i) \right\rangle, \label{eq:Mellin_n_subleading_soft_photon_thm}
\een
for $j=1/2, 1, 3/2,\cdots$. In the above equations we have kept the helicity and all other internal quantum numbers of hard particles implicit because the soft factors depend only on the electric charges and the momenta of them. 
The soft operator $\mathcal{S}^\Delta(w,\bar{w})$ for positive helicity outgoing soft photon has been defined via the Mellin transformation of the photon annihilation operator $a(k, \sigma=+1)$ along with the conformal soft limit \cite{Donnay:2018neh} taken inside the celestial correlation functions as\footnote{Here we used the following result of Mellin transformation $\int_0^\infty d\omega\, \omega^{s-1} \omega^n (\ln\omega)^k=\f{(-1)^k k!}{(s+n)^{k+1}}$, as derived in \cite{Krishna:2023ukw,Guevara:2019ypd}. One may be tempted to resum all the multi-order poles in the scaling dimensions and interpret the resummed Mellin-space Ward identity as a simple pole in $\Delta$ with anomalous dimensions. However, the resulting structure of the sum appears to be of the form $\sum_{n=0}^\infty (\Delta + n)^{n+2}$ in Mellin space, or equivalently exhibits the integral behavior $\int_0^\infty d\omega \omega^{\omega + \Delta - 2}$, which falls into the category of the ``Sophomore’s Dream'', for which no closed analytical result is known.}
\ben
\begin{gathered}
\mathcal{S}^{1-2j}(w,\bar{w}) = -\sqrt{2}(4\pi^2)^{2j} \f{(-1)^{2j}}{(2j)!}\lim_{\Delta\rightarrow 1-2j} (\Delta +2j-1 )^{2j+1} \Gamma_{\Delta}^+(w,\bar{w}),\\
\Gamma_{\Delta}^+(w,\bar{w})= \int_0^\infty d\omega \, \omega^{\Delta-1}\ a(-\omega\, \mathbf{n}(w,\bar{w}), \sigma = +1),
\end{gathered}
\een
for $j=0,1/2, 1, 3/2,\cdots$. The holomorphic and anti-holomorphic conformal dimensions of $\mathcal{S}^{1-2j}$ are $h=1-j,\, \bar h=-j$ as follows from the (Lorentz) conformal invariance of \eqref{eq:Mellin_leading_soft_photon_thm} and \eqref{eq:Mellin_n_subleading_soft_photon_thm}.

The Weinberg soft theorem \eqref{eq:Mellin_leading_soft_photon_thm} is known \cite{Strominger:2017zoo,Strominger:2013lka} to be equivalent to the Ward-identity of a $U(1)$ Kac-Moody algebra at level zero. In the rest of the paper we focus on the symmetry interpretation of the all loop soft theorems \eqref{eq:Mellin_n_subleading_soft_photon_thm}.


\section{$\mathcal{O}(\ln\omega)$ soft theorem as Ward identity and the dipole current operator}\label{Dipole}

In the conformal primary basis, the $\mathcal{O}(\ln\omega)$ soft theorem \eqref{eq:Mellin_n_subleading_soft_photon_thm} $(j=1/2)$ can be written as 

\be
\left\langle \mathcal{S}^0(w,\bar{w}) \prod_{i=1}^N \phi^{(\eta_i)}_{\D_i}(z_i,\bar{z}_i,e_i) \right\rangle
= \sum_{a=1}^N \f{\eta_a {e_a}^2}{w-z_a}\Bigg[ \sum_{\substack{b=1\\ b\neq a}}^N e_b\,  \f{\bar w-\bar z_b}{\bar z_a-\bar z_b}\Bigg]
\left\langle \phi^{(\eta_a)}_{\D_a-1}(z_a,\bar{z}_a,e_a) \prod_{\substack{i=1\\ i\neq a}}^N \phi^{(\eta_i)}_{\D_i}(z_i,\bar{z}_i,e_i) \right\rangle .
\ee

This equation, as it stands, does not allow us to define a Celestial OPE between soft and hard operators because of the multi-particle sum. For example, if we take $w\rightarrow z_a$ then the most singular term in $(w-z_a)$ 
\be\label{nope}
\mathcal{S}^0(w,\bar w) \phi^{(\eta_a)}_{\D_a}(z_a, \bar z_a, e_a) \sim \frac{\eta_a e_a^2}{w-z_a} \( \sum_{\substack{b=1\\ b\neq a}}^N e_b \frac{\bar w - \bar z_b}{\bar z_a - \bar z_b}\) \phi^{(\eta_a)}_{\D_a-1}(z_a, \bar z_a, e_a),
\ee
does not have the standard form of an OPE. To cure this problem we introduce the ``dipole current operator" $\bar d(\bar{w},\bar{z})$ whose correlation function is given by
\be
\boxed{
\left\langle \bar d(\bar{w},\bar{z}) \prod_{i} \phi^{(\eta_i)}_{\D_i}(z_i,\bar{z}_i,e_i) \right\rangle= \sum_a e_a \f{\bar{w}-\bar{z}_a}{\bar{z}-\bar{z}_a} \left\langle\prod_{i} \phi^{(\eta_i)}_{\D_i}(z_i,\bar{z}_i,e_i) \right\rangle .}\label{eq:dipole_op_corr}
\ee

From \eqref{eq:dipole_op_corr} it is evident that the charge $Q_{\bar d}(\bar w)$ defined through $Q_{\bar d}(\bar{w})= \oint \f{d\bar{z}}{2\pi i} \bar d(\bar{w},\bar{z})$ measures the (anti-holomorphic) dipole moment of a charged operator, i.e, 
\ben
\[Q_{\bar d}(\bar{w}), \phi^{(\eta)}_{\D}(z,\bar{z},e)\]
&&= - e(\bar z - \bar w) \phi^{(\eta)}_{\D}(z,\bar{z},e),
\een
justifying the name ``dipole current operator". 

Now the $\mathcal{O}(\ln\omega)$ soft theorem can be rewritten as a Ward identity
\be\label{eq:Subleading_Ward_identity}
\boxed{
\left\langle \mathcal{S}^0(w,\bar{w}) \prod_{i=1}^N \phi^{(\eta_i)}_{\D_i}(z_i,\bar{z}_i,e_i) \right\rangle
= \sum_{a=1}^N \f{\eta_ae_a^2}{w-z_a}\left\langle :\bar d\phi^{(\eta_a)}_{\D_a -1}:(\bar{w},z_a,\bar{z}_a,e_a) \prod_{\substack{i=1\\ i\neq a}}^N \phi^{(\eta_i)}_{\D_i}(z_i,\bar{z}_i,e_i) \right\rangle,}
\ee
where the composite operator $:\bar d\phi^{(\eta)}_{\D}:(\bar w, z, \bar z, e)$ is defined in the standard way as 
\ben\label{eq:D_normal_ordering}
\begin{gathered}
:\bar d \phi^{(\eta)}_{\D}:(\bar{w},z,\bar{z},e)\equiv\lim\limits_{\bar{\xi}\rightarrow \bar{z}} \Bigg(\bar d(\bar{w},\bar{\xi})  \phi^{(\eta)}_{\D}(z,\bar{z},e)
 - e \f{\bar{w}-\bar{z}}{\bar{\xi}-\bar{z}} \phi^{(\eta)}_{\D}(z,\bar{z},e)\Bigg).
 \end{gathered}
\een
The Ward identity \eqref{eq:Subleading_Ward_identity} allows us to write down the Celestial OPE between the soft operator $\mathcal{S}^0$ and a hard operator as 
\be
\boxed{
\mathcal{S}^0(w,\bar w) \phi^{(\eta)}_{\D}(z,\bar z, e) \sim \frac{\eta e^2}{w-z} :\bar d\phi^{(\eta)}_{\D-1}:(\bar w, z,\bar z, e)\, .}
\ee
The equation \eqref{eq:dipole_op_corr} implies that: \\
The dual Celestial CFT$_2$ has a ``Dipole Symmetry" generated by the current $\bar d(\bar w,\bar z)$. The existence of the dipole current follows from the locality of the celestial CFT$_2$ in the presence of loop corrections in the bulk. 

 \section{Conformal transformation law of the dipole operator}\label{S:algebra}
 Under the $SL(2,\mathbb{R})_{\text{R}}$ transformation $\bar{z}\rightarrow \bar{z}'=\f{a\bar{z}+b}{c\bar{z}+d}$ with $a,b,c,d\in \mathbb{R}$ and $ad-bc=1$,  the invariance of the correlation function  \eqref{eq:dipole_op_corr} requires the dipole current $\bar d(\bar{w},\bar{z})$ to transform in the following way,
\ben\label{finitetr}
\bar d(\bar w, \bar z)\rightarrow\bar d'(\bar{w},\bar{z})=\f{c\bar w+d}{c\bar z+d}\times\bar d\left(\bar{w}',\bar{z}'\right).\label{eq:dipole_transformation}
\een
Note that the RHS of the correlation function \eqref{eq:dipole_op_corr} is a polynomial of degree one in $\bar w$. So we can Taylor expand $\bar d(\bar w, \bar z)$ in $\bar w$ coordinate \cite{Banerjee:2020zlg,Guevara:2021abz,Strominger:2021mtt} and write 
\be\label{doublet}
\bar d(\bar w, \bar z) = \bar{d}^{\f{1}{2}}_{-\frac{1}{2}}(\bar z) + \bar w \  \bar d^{\f{1}{2}}_{\frac{1}{2}}(\bar z).
\ee
In terms of the doublet $(\bar d^{\frac{1}{2}}_{\frac{1}{2}},\bar d^{\frac{1}{2}}_{-\frac{1}{2}})$ the transformation \eqref{eq:dipole_transformation} becomes
\ben\label{trdoublet}
\bar d^{\prime\f{1}{2}}_\f{1}{2}(\bar z)=\f{1}{c\bar z+d}\left(a\, \bar d^{\f{1}{2}}_\f{1}{2}(\bar z')+c\, \bar d^{\f{1}{2}}_{-\f{1}{2}}(\bar z') \right),\quad \bar d^{\prime\f{1}{2}}_{-\f{1}{2}}(\bar z)=\f{1}{c\bar z+d}\left(b\, \bar d^{\f{1}{2}}_\f{1}{2}(\bar z')+d\, \bar d^{\f{1}{2}}_{-\f{1}{2}}(\bar z') \right).
\een
and for infinitesimal translation, dilatation and special conformal transformations (SCT)\footnote{
\ben
\text{Translation:} \quad \begin{pmatrix}
1 & b \\
0 & 1 \end{pmatrix}, \qquad  \text{Dilatation:} \quad \begin{pmatrix}
\lambda^\f{1}{2} & 0 \\
0 & \lambda^{-\f{1}{2}} \end{pmatrix}, \qquad  \text{SCT:} \quad \begin{pmatrix}
1 & 0 \\
-c & 1 \end{pmatrix}, 
\een}
we get the following commutation relations
\ben\label{eq:SL2R_commutation}
\begin{gathered}
\left[\bar L_{-1},\bar d^j_m(\bar z)\right]=\bar\p \bar d^j_m(\bar z)+(j-m)\bar d^j_{m+1},\\
\left[\bar L_{0},\bar d^j_m(\bar z)\right]=\left(j+m+\bar z\bar\p\right)\bar d^j_m(\bar z),\\
\left[\bar L_{1},\bar d^j_m(\bar z)\right]=\left(2j\bar z+\bar z^2\bar\p\right)\bar d^j_m(\bar z) -(j+m)\bar d^j_{m-1}(\bar z).
\end{gathered}
\een
Here $j=\frac{1}{2}$, $m = \pm\frac{1}{2}$ and the generators of translation, dilatation and SCT are denoted by $\bar L_{-1},\, \bar{L}_0$ and $\bar{L}_1$ respectively. 

The transformation law \eqref{trdoublet} shows that the doublet $\left(\bar{d}^{\f{1}{2}}_{\frac{1}{2}}(\bar z), \bar{d}^{\f{1}{2}}_{-\frac{1}{2}}(\bar z)\right)$ transform in the tensor product of two $SL(2,\mathbb{R})_{\text{R}}$ representations one of which is a conformal highest weight representation with weight $\bar h=1/2$ and the other one is a two dimensional spin-$\frac{1}{2}$ representation. Under the spin-$\frac{1}{2}$ representation the fields $\bar{d}^{\f{1}{2}}_{\pm\frac{1}{2}}(\bar z)$ have weights $\pm\frac{1}{2}$. To see this at the level of infinitesimal transformation law, we denote the generator in the highest weight ($\bar h= \frac{1}{2}$) representation by $\bar L_n^{\bar h}$ and the generator in the spin-$\f{1}{2}$ representation by $\bar L_n^{\bar \sigma}$. Now one can easily verify that the following commutation relations 
\ben\label{eq:SL2R_commutation_split}
\begin{gathered}
\left[\bar L^{\bar h}_n,\bar d^j_m(\bar z)\right]=\left(\bar z^{n+1}\bar\p +(n+1)j\bar z^n\right)\bar d^j_m(\bar z), \\
\left[\bar L^{\bar{\sigma}}_{-1},\bar d^j_m(\bar z)\right]=(j-m)\, \bar d^j_{m+1}(\bar z),\\
\left[\bar L^{\bar{\sigma}}_{0},\bar d^j_m(\bar z)\right]=m\, \bar d^j_m(\bar z),\\
\left[\bar L^{\bar{\sigma}}_{1},\bar d^j_m(\bar z)\right]= -(j+m)\, \bar d^j_{m-1}(\bar z),
\end{gathered}
\een 
reproduce the transformation law \eqref{eq:SL2R_commutation} for $n=-1,0,1$,   $j = \frac{1}{2}$ and  $m = \pm \frac{1}{2}$ provided the relation $\bar L_n=\bar L_n^{\bar h}+\bar L_n^{\bar \sigma}$ holds.
 

 
\section{Some notation}\label{S:notation}
 To make the $SL(2,\mathbb{R})_{\text{R}}$ covariance manifest it will be useful to introduce the two-component spinor notation. We define 
\be
\bar Z^\alpha = 
\begin{pmatrix}
\bar z \\
1
\end{pmatrix} , \ 
\epsilon_{\alpha\beta} =
\begin{pmatrix}
0 & 1 \\
-1& 0
\end{pmatrix}, \ 
\bar Z_\alpha = \epsilon_{\alpha\beta} \bar Z^\beta = 
\begin{pmatrix}
1 \\
-\bar z
\end{pmatrix}, \  \alpha,\beta = \pm 1/2,
\ee
 so that \eqref{doublet} can be rewritten as 
\be
\bar d(\bar w, \bar z) = \bar W^\alpha \bar d^{\f{1}{2}}_\alpha(\bar z),
\ee
and the Ward identity \eqref{eq:dipole_op_corr} for the dipole symmetry becomes\footnote{Note that $\bar z - \bar z_a = \bar Z^\alpha \bar Z_{a\alpha}$, how ever we are keeping it explicit.}
\be
\left\langle \bar d^{\f{1}{2}}_{\alpha}(\bar z) \prod_{i} \phi^{(\eta_i)}_{\D_i}(z_i,\bar{z}_i,e_i) \right\rangle= \sum_a e_a \f{\bar Z_{a\alpha}}{\bar{z}-\bar{z}_a} \left\langle\prod_{i} \phi^{(\eta_i)}_{\D_i}(z_i,\bar{z}_i,e_i) \right\rangle .
\ee

We can also expand the soft photon operator $\mathcal S^0(w,\bar w)$ in a similar way 
\be
\mathcal S^0(w,\bar w) = \bar W^\alpha \mathcal S^{0}_\alpha(w),
\ee 
and rewrite the $\mathcal{O}(\ln\omega)$ soft theorem as 
\ben
\begin{gathered}
\left\langle \mathcal{S}^0_{\alpha}(w) \prod_{i=1}^N \phi^{(\eta_i)}_{\D_i}(z_i,\bar{z}_i,e_i) \right\rangle
= \sum_{a=1}^N \f{\eta_a {e_a}^2}{w-z_a}
\left\langle :\bar d^{\f{1}{2}}_{\alpha}\phi^{(\eta_a)}_{\D_a-1}:(z_a,\bar{z}_a,e_a) \prod_{\substack{i=1\\ i\neq a}}^N \phi^{(\eta_i)}_{\D_i}(z_i,\bar{z}_i,e_i) \right\rangle .
\end{gathered}
\een

\section{Correlation functions of dipole currents and matter fields}
The dipole operator $\bar d^{\f{1}{2}}_\alpha$ has scaling dimension $1/2$. Therefore conformal invariance determines the two point function to be \footnote{In terms of the original field $\bar d(\bar w, \bar z)$ this is nothing but $\langle \bar d(\bar w_1,\bar z_1) \bar d(\bar w_2, \bar z_2)\rangle = k \frac{\bar w_1 - \bar w_2}{\bar z_1 - \bar z_2}$. This form of the two point function is uniquely determined by the $SL(2,\mathbb{R})_{\text{R}}$ transformation law \eqref{finitetr}. }
\be\label{eq:dd_corr}
\langle\bar d^{\f{1}{2}}_{\alpha}(\bar z_1) \bar d^{\f{1}{2}}_{\beta}(\bar z_2)\rangle = k \frac{\epsilon_{\alpha\beta}}{\bar z_1 - \bar z_2}\, ,
\ee
where $k$ is a real number which can be \textit{zero} also. However, as we will see we can take $k$ to be \textit{nonzero} \textit{without changing the single soft photon theorems}. Our current understanding of the role of dipole symmetry in the loop level soft theorems does not allow us to determine the value of $k$. This is discussed in more detail in section \ref{kk}. 

The singular terms in the OPE between two dipole operators is given by 
\be\label{eq:dd_OPE}
\bar d^{\f{1}{2}}_\alpha(\bar z_1) \bar d^{\f{1}{2}}_\beta({\bar z_2}) \sim k \frac{\epsilon_{\alpha\beta}}{\bar z_1 - \bar z_2}\, .
\ee
This together with the boundary condition that in the absence of any operator insertion at infinity $\bar d^{\f{1}{2}}_\alpha(\bar z) \rightarrow 1/\bar z$ as $\bar z\rightarrow\infty$ determine the correlation function with multiple insertion of dipole currents to be
\ben
\begin{gathered}
\left\langle \bar d^{\f{1}{2}}_{\alpha_1}(\bar\xi_{1})\prod_{p=2}^n\bar d^{\f{1}{2}}_{\alpha_p}(\bar{\xi_p}) \prod_{i} \phi^{(\eta_i)}_{\D_i}(z_i,\bar{z}_i,e_i) \right\rangle= \(\sum_a e_a \f{\bar Z_{a{\alpha_1}}}{\bar{\xi_1}-\bar{z}_a}\) \left\langle\prod_{p=2}^n\bar d^{\f{1}{2}}_{\alpha_p}(\bar{\xi_p})\prod_{i} \phi^{(\eta_i)}_{\D_i}(z_i,\bar{z}_i,e_i) \right\rangle \\
+ \sum_{a=2}^n k \frac{\epsilon_{\alpha_1\alpha_a}}{\bar\xi_1 - \bar\xi_a} \left\langle\bar d^{\f{1}{2}}_{\alpha_2}(\xi_2)\cdots \cancel{\bar d^{\f{1}{2}}_{\alpha_a}(\bar\xi_a)}\cdots \bar d^{\f{1}{2}}_{\alpha_n}(\bar\xi_n)\prod_{i} \phi^{(\eta_i)}_{\D_i}(z_i,\bar{z}_i,e_i) \right\rangle .
\end{gathered}
\een
This correlation function allows us to define currents of different weights and spins and play an important role in this paper. For example, let us consider the correlation function 
\ben
\begin{gathered}
\left\langle \bar d^{\f{1}{2}}_{\alpha_1}(\bar\xi_{1}) \bar d^{\f{1}{2}}_{\alpha_2}(\bar\xi_{2}) \prod_{i} \phi^{(\eta_i)}_{\D_i}(z_i,\bar{z}_i,e_i) \right\rangle= \\ \(\sum_a e_a \f{\bar Z_{a{\alpha_1}}}{\bar{\xi_1}-\bar{z}_a}\) \(\sum_b e_b \f{\bar Z_{b{\alpha_2}}}{\bar{\xi_2}-\bar{z}_b}\) \left\langle \prod_{i} \phi^{(\eta_i)}_{\D_i}(z_i,\bar{z}_i,e_i) \right\rangle 
+ k \frac{\epsilon_{\alpha_1\alpha_2}}{\bar\xi_1 - \bar\xi_2} \left\langle\prod_{i} \phi^{(\eta_i)}_{\D_i}(z_i,\bar{z}_i,e_i) \right\rangle,
\end{gathered}
\een
with two insertions of the dipole current. This allows us to define a dimension $\bar h=1$ and spin-1 current $\bar d^1_{\alpha\beta}$ as 
\be\label{eq:d1_def}
\bar d^1_{\alpha\beta}(\bar z) = \bar d^1_{\beta\alpha}(\bar z) = \lim_{\bar z' \rightarrow \bar z} \( \bar d^{\f{1}{2}}_{\alpha}(\bar z') \bar d^{\f{1}{2}}_{\beta}(\bar z) - k \frac{\epsilon_{\alpha\beta}}{\bar z' - \bar z }\) = : \bar d^{\f{1}{2}}_\alpha(\bar z)\bar d^{\f{1}{2}}_\beta(\bar z):\, ,
\ee
whose correlation function with matter is given by 
\ben\label{spin1}
\begin{gathered}
\left\langle \bar d^{1}_{\alpha_1\alpha_2}(\bar\xi) \prod_{i} \phi^{(\eta_i)}_{\D_i}(z_i,\bar{z}_i,e_i) \right\rangle=  \(\sum_a e_a \f{\bar Z_{a{\alpha_1}}}{\bar{\xi}-\bar{z}_a}\) \(\sum_b e_b \f{\bar Z_{b{\alpha_2}}}{\bar{\xi}-\bar{z}_b}\) \left\langle \prod_{i} \phi^{(\eta_i)}_{\D_i}(z_i,\bar{z}_i,e_i) \right\rangle .
\end{gathered}
\een
In a similar fashion we can define spin-$j$ current $\bar d^j_{\alpha_1\cdots \alpha_{2j}}$ by generalizing the normal ordering \eqref{eq:d1_def} for $2j$ number of $\bar d^\f{1}{2}$ in a standard way by removing all the singular terms and get
\be
\bar d^j_{\alpha_1\cdots \alpha_{2j}} = : \bar d^{\f{1}{2}}_{\alpha_1} \bar d^{\f{1}{2}}_{\alpha_2}\cdots \bar d^{\f{1}{2}}_{\alpha_{2j}} :\, ,\label{eq:d_j_def}
\ee
whose correlation function with matter is given by 
\ben\label{spinj}
\begin{gathered}
\left\langle \bar d^{j}_{\alpha_1\alpha_2\cdots\alpha_{2j}}(\bar\xi) \prod_{i} \phi^{(\eta_i)}_{\D_i}(z_i,\bar{z}_i,e_i) \right\rangle=  \prod_{k=1}^{2j}\left(\sum_a e_a \f{\bar Z_{a{\alpha_k}}}{\bar{\xi}-\bar{z}_a}\right)  \left\langle \prod_{i} \phi^{(\eta_i)}_{\D_i}(z_i,\bar{z}_i,e_i) \right\rangle .
\end{gathered}
\een

\section{$\mathcal{O}(\omega^{2j-1} (\ln\omega)^{2j})$ soft theorem and Celestial OPE}
In this section, we show how to write down the OPE between the soft photon and a hard charged operator using the soft theorems in  \eqref{eq:Mellin_n_subleading_soft_photon_thm}. As $w\rightarrow z_a$ the Weinberg \eqref{eq:Mellin_leading_soft_photon_thm} and the singular parts of the logarithmic \eqref{eq:Mellin_n_subleading_soft_photon_thm} soft theorems can be combined to give
\ben\label{eq:Ward_identity_singular}
\begin{gathered}
\left\langle \mathcal{S}^{1-2j}(w,\bar{w}) \prod_{i=1}^N \phi^{(\eta_i)}_{\D_i}(z_i,\bar{z}_i,e_i) \right\rangle_{w\rightarrow z_a}\\
= \f{1}{(2j)!}\f{\eta_a^{2j}e_a^{2j+1}}{w-z_a}\Bigg[ \sum_{\substack{b=1\\ b\neq a}}^N  e_b\,  \f{\bar w-\bar z_b}{\bar z_a-\bar z_b}\Bigg]^{2j}\left\langle \phi^{(\eta_a)}_{\D_a - 2j}(z_a,\bar{z}_a,e_a) \prod_{\substack{i=1\\ i\neq a}}^N \phi^{(\eta_i)}_{\D_i}(z_i,\bar{z}_i,e_i) \right\rangle + \text{non-singular terms}
\end{gathered}
\een
where $j=0,\f{1}{2},1,\cdots$. Let us first discuss the $j=1$ case. 


We write the $j=1$ soft photon $\mathcal{S}^{-1}(w,\bar w)$ as
\be
\mathcal{S}^{-1}(w,\bar w) = \frac{1}{2}\bar W^{\alpha} \bar W^\beta \mathcal{S}^{-1}_{\alpha\beta}(w),
\ee
and rewrite \eqref{eq:Ward_identity_singular} as
\be\label{rewrite}
\begin{gathered}
\left\langle \mathcal{S}^{-1}_{\alpha\beta}(w) \prod_{i=1}^N \phi^{(\eta_i)}_{\D_i}(z_i,\bar{z}_i,e_i) \right\rangle_{w\rightarrow z_a}\\
=  \f{\eta_a^{2} e_a^{3}}{w-z_a}\( \sum_{\substack{b=1\\ b\neq a}}^N  e_b\,  \f{\bar Z_{b\alpha}}{\bar z_a-\bar z_b}\)\( \sum_{\substack{c=1\\ c\neq a}}^N  e_c\,  \f{\bar Z_{c\beta}}{\bar z_a-\bar z_c}\)  \left\langle \phi^{(\eta_a)}_{\D_a - 2}(z_a,\bar{z}_a,e_a) \prod_{\substack{i=1\\ i\neq a}}^N \phi^{(\eta_i)}_{\D_i}(z_i,\bar{z}_i,e_i) \right\rangle \\ + \text{non-singular terms}
\end{gathered}
\ee
Now by comparing with \eqref{spin1}, we write \eqref{rewrite} as
\be
\begin{gathered}
\left\langle \mathcal{S}^{-1}_{\alpha\beta}(w) \prod_{i=1}^N \phi^{(\eta_i)}_{\D_i}(z_i,\bar{z}_i,e_i) \right\rangle_{w\rightarrow z_a} = \f{\eta_a^{2} e_a^{3}}{w-z_a} \left\langle :\bar d^1_{\alpha\beta}\phi^{(\eta_a)}_{\D_a - 2}:(z_a,\bar{z}_a,e_a) \prod_{\substack{i=1\\ i\neq a}}^N \phi^{(\eta_i)}_{\D_i}(z_i,\bar{z}_i,e_i) \right\rangle \\ + \text{non-singular terms}
\end{gathered}
\ee
with the normal ordering definition
\ben\label{eq:d1_phi}
\begin{gathered}
:\bar d^1_{\alpha\beta}\phi^{(\eta)}_{\D}(z,\bar z, e): \quad =\quad  \lim_{\bar z'\rightarrow \bar z} \Big[ \bar d^1_{\alpha\beta}(\bar z')\phi^{(\eta)}_{\D}(z,\bar z, e) - e^2 \frac{\bar Z_\alpha \bar Z_\beta}{(\bar z' - \bar z)^2} \phi^{(\eta)}_{\D}(z,\bar z, e)\\
 - e \frac{\bar Z_\alpha}{\bar z' - \bar z} :\bar d^{\f{1}{2}}_{\beta}\phi^{(\eta)}_{\D}:(z,\bar z, e)  -   e \frac{\bar Z_\beta}{\bar z' - \bar z} :\bar d^{\f{1}{2}}_{\alpha}\phi^{(\eta)}_{\D}:(z,\bar z, e)   \Big].
\end{gathered}
\een

Now generalization of the above construction to arbitrary values of $j$ is simple. We write the soft operator in spinor notation
\be
\mathcal{S}^{1-2j}(w,\bar w) = \frac{1}{(2j)!}\bar W^{\alpha_1} \bar W^{\alpha_2}\cdots \bar W^{\alpha_{2j}} \mathcal{S}^{1-2j}_{\alpha_1\alpha_2\cdots\alpha_{2j}}(w),
\ee
and \eqref{eq:Ward_identity_singular} can be rewritten as
\be\label{eq:singular_Ward_identity}
\begin{gathered}
\left\langle \mathcal{S}^{1-2j}_{\alpha_1\alpha_2\cdots\alpha_{2j}}(w) \prod_{i=1}^N \phi^{(\eta_i)}_{\D_i}(z_i,\bar{z}_i,e_i) \right\rangle_{w\rightarrow z_a}\\
=  \f{\eta_a^{2j} e_a^{2j+1}}{w-z_a}\left\langle :\bar d^{j}_{\alpha_1\alpha_2\cdots\alpha_{2j}}\phi^{(\eta_a)}_{\D_a - 2j}:(z_a,\bar{z}_a,e_a) \prod_{\substack{i=1\\ i\neq a}}^N \phi^{(\eta_i)}_{\D_i}(z_i,\bar{z}_i,e_i) \right\rangle + \text{non-singular terms}
\end{gathered}
\ee
where the definition of the higher spin current $\bar d^j_{\alpha_1\alpha_2\cdots\alpha_{2j}}(\bar z)$ is given in \eqref{eq:d_j_def}. The equation \eqref{eq:Ward_identity_singular} leads to the following OPE between the soft operator $\mathcal{S}^{1-2j}_{\alpha_1\alpha_2\cdots\alpha_{2j}}(w)$ and a hard operator $\phi^{(\eta_a)}_{\D_a}(z_a,\bar z_a, e_a)$ 
\be\label{hope}
\mathcal{S}^{1-2j}_{\alpha_1\alpha_2\cdots\alpha_{2j}}(w) \phi^{(\eta_a)}_{\D_a}(z_a,\bar z_a, e_a) = \f{\eta_a^{2j} e_a^{2j+1}}{w-z_a} :\bar d^{j}_{\alpha_1\alpha_2\cdots\alpha_{2j}}\phi^{(\eta_a)}_{\D_a - 2j}:(z_a,\bar{z}_a,e_a) + \text{non-singular terms}
\ee
The current $\bar d^j_{\alpha_1\alpha_2\cdots \alpha_{2j}}$ is completely symmetric in the spinor indices and transforms in the tensor product of two $SL(2,\mathbb{R})_{\text{R}}$ representations one of which is a conformal highest weight representation of weight $\bar h=j$ and the other one is the $(2j+1)-$dimensional spin-$j$ representation. We discuss this in more detail in section \ref{confhs}.

Now unlike the $j=0$ and $j=\frac{1}{2}$ cases, the higher $j$ soft theorems cannot be written as Ward identities using the higher spin currents alone. However, as equation \eqref{hope} shows the singular term in the OPE is completely determined in terms of the higher spin current. To write the soft theorems as Ward identities for $j\ge 1$ we need to understand the part of the soft theorem which does not have a pole in $w$. We leave this interesting question for future investigation.

In the following section we discuss the algebra of the higher spin currents for general values of $k$. 

\section{Algebra of higher spin currents}
Since the contraction between two dipole currents $\bar d^\f{1}{2}$ is a $c$-number we can compute the OPE between the currents defined in \eqref{eq:d_j_def} using Wick's theorem. We show some results for the singular terms only: 
\be
\bar d^{1/2}_{\alpha_1}(\bar z_1) \bar d^{1/2}_{\alpha_2}(\bar z_2) \sim k \frac{\epsilon_{\alpha_1\alpha_2}}{\bar z_1 - \bar z_2}\, ,
\ee
\be
\bar d^{1}_{\alpha_1\alpha_2}(\bar z_1) \bar d^{1/2}_{\alpha_3}(\bar z_2) \sim k \frac{\epsilon_{\alpha_1\alpha_3} \bar d^{1/2}_{\alpha_2}(\bar z_2) + \epsilon_{\alpha_2\alpha_3} \bar d^{1/2}_{\alpha_1}(\bar z_2)}{\bar z_1 - \bar z_2}\, ,
\ee
\ben
\begin{gathered}
\bar d^{1}_{\alpha_1\alpha_2}(\bar z_1) \bar d^{1}_{\alpha_3\alpha_4}(\bar z_2) \sim k \frac{\epsilon_{\alpha_1\alpha_3}\bar d^1_{\alpha_2\alpha_4}(\bar z_2 ) + \epsilon_{\alpha_1\alpha_4}\bar d^1_{\alpha_2\alpha_3}(\bar z_2) + \epsilon_{\alpha_2\alpha_4}\bar d^1_{\alpha_1\alpha_3}(\bar z_2)+ \epsilon_{\alpha_2\alpha_3}\bar d^1_{\alpha_1\alpha_4}(\bar z_2)}{\bar z_1 -\bar z_2} \\
+ k^2 \frac{\epsilon_{\alpha_1\alpha_3}\epsilon_{\alpha_2\alpha_4} + \epsilon_{\alpha_1\alpha_4}\epsilon_{\alpha_2\alpha_3}}{(\bar z_1 -\bar z_2)^2}\, ,
\end{gathered}
\een
\ben\label{break}
\begin{gathered}
\bar d^\f{3}{2}_{\alpha_1\alpha_2\alpha_3}(\bar z_1)\, \bar d^\f{3}{2}_{\alpha_4\alpha_5\alpha_6}(\bar z_2) \\
\sim \f{1}{\bar z_1-\bar z_2}\Bigg(\epsilon_{\alpha_1\alpha_4}\Big(k \ \bar d^2_{\alpha_2\alpha_3\alpha_5\alpha_6}(\bar z_2)+\textcolor{blue}{\f{k^2}{2} \epsilon_{\alpha_2\alpha_5}:\bar\p \bar d^\f{1}{2}_{\alpha_3}(\bar z_2)\bar d^\f{1}{2}_{\alpha_6}(\bar z_2)}:\\
\textcolor{blue}{+\f{k^2}{2} \epsilon_{\alpha_3\alpha_5}:\bar\p \bar d^\f{1}{2}_{\alpha_2}(\bar z_2)\bar d^\f{1}{2}_{\alpha_6}(\bar z_2):+\f{k^2}{2} \epsilon_{\alpha_2\alpha_6}:\bar\p \bar d^\f{1}{2}_{\alpha_3}(\bar z_2)d^\f{1}{2}_{\alpha_5}(\bar z_2):+\f{k^2}{2} \epsilon_{\alpha_3\alpha_6}:\p \bar d^\f{1}{2}_{\alpha_2}(\bar z_2)\bar d^\f{1}{2}_{\alpha_5}(\bar z_2)}:\Big)\\
+ \text{ 8 terms under permutation}\Bigg) \\
+ k^2 \Bigg( \f{\epsilon_{\alpha_1\alpha_4}\epsilon_{\alpha_2\alpha_5}+\epsilon_{\alpha_1\alpha_5}\epsilon_{\alpha_2\alpha_4}}{(\bar z_1-\bar z_2)^2}\bar d^1_{\alpha_3\alpha_6}(\bar z_2)+ \text{ 8 terms under permutation}\Bigg)\\
+ k^3 \ \f{\epsilon_{\alpha_1\alpha_4}\epsilon_{\alpha_2\alpha_5}\epsilon_{\alpha_3\alpha_6}+\text{5 terms under permutation}}{(\bar z_1-\bar z_2)^3}\, . 
\end{gathered}
\een
We can see that the currents $\bar d^1_{\alpha\beta}$ generate a $SL(2,\mathbb{R})$ Kac-Moody algebra. The currents $\bar d^{1/2}_{\alpha}$ and $\bar d^1_{\alpha\beta}$ form a closed subalgebra. However, starting from spin-$\frac{3}{2}$ the algebra of currents do not close among themselves. This is evident from the existence of the blue coloured terms in the OPE \eqref{break}. To understand the underlying algebra better let us take the semiclassical limit of $k(\sim i\hbar)\rightarrow 0$. In this limit the problematic terms which are all of $\mathcal{O}(k^2)$ and higher are subleading. The leading terms of $\mathcal{O}(k)$ in the OPE are those which arise from single contractions. If we keep only the leading terms then we get the closed OPE
\be\label{closed}
\begin{gathered}
\bar d^{j_1}_{\alpha_1\cdots\alpha_{2j_1}}(\bar z_1) \bar d^{j_2}_{\beta_1\cdots\beta_{2j_2}}(\bar z_2) \sim 
k \ \frac{\epsilon_{\alpha_1\beta_1}\bar d^{j_1 + j_2 -1}_{\alpha_2\cdots\alpha_{2j_1}\beta_2\cdots\beta_{2j_2}}(\bar z_2) + \text{Contractions}}{\bar z_1 - \bar z_2} + \mathcal{O}(k^2).
\end{gathered}
\ee
Now we show that the global part of the algebra generated by the higher spin currents in the semiclassical limit is the wedge subalgebra of the $w_{1+\infty}$ algebra. We denote the wedge subalgebra by $\wedge w_{1+\infty}$.

We define the global modes
\be
\bar{\mathbf{d}}^j_{\alpha_1\alpha_2\cdots \alpha_{2j}} =  \oint \f{d\bar z}{2\pi i} \ \bar d^j_{\alpha_1\alpha_2\cdots\alpha_{2j}}(\bar z).
\ee
The algebra of these global modes is given by 
\be\label{global}
\begin{gathered}
\left[ \bar{\mathbf{d}}^{j_1}_{\alpha_1\alpha_2\cdots \alpha_{2j_1}}, \bar{\mathbf{d}}^{j_2}_{\beta_1\beta_2\cdots \beta_{2j_2}} \right] = k \  \( \epsilon_{\alpha_1\beta_1}\bar{\mathbf{d}}^{j_1 + j_2 -1}_{\alpha_2\cdots\alpha_{2j_1}\beta_2\cdots\beta_{2j_2}} + \cdots + \epsilon_{\alpha_{2j_1}\beta_{2j_2}}\bar{\mathbf{d}}^{j_1 + j_2 -1}_{\alpha_1\cdots\alpha_{2j_1-1}\beta_1\cdots\beta_{2j_2-1}}\),
\end{gathered}
\ee
containing $2j_1\times 2j_2$ number of terms in the RHS. 
To show that this algebra is isomorphic to $\wedge w_{1+\infty}$ let us define 
\be
w^p_n \equiv \frac{1}{2} \bar{\mathbf{d}}^j_{\alpha_1\alpha_2\cdots \alpha_{2j}} ,
\ee
where 
\be\label{w}
p = j+1, \  n = \sum_{i=1}^{2j} \alpha_i, \ \alpha_i = \pm \f{1}{2}, \  j\in \frac{1}{2}\mathbb{Z_{+}}\, .
\ee
It follows from \eqref{w} that the maximum value of $n$ is $j = p-1$ and the minimum value of $n$ is $-j = 1-p$. In other words 
\be\label{restriction}
1-p \le n \le p-1.
\ee
It is easy to show that in terms of the generators $w^p_n$ the algebra \eqref{global} can be rewritten as 
\be\label{wedge}
\left[w^p_m, w^q_n\right] = k \left[ m(q-1) - n(p-1) \right] w^{p+q-2}_{m+n},
\ee
with $p,q=\f{3}{2},2,\f{5}{2},3,\cdots$ and the value for $w^1_0=\f{1}{2}$ appears when $p=q=\f{3}{2}$.

The algebra \eqref{wedge} subject to the restriction \eqref{restriction} is well known to be the wedge subalgebra $\wedge w_{1+\infty}$ \cite{Bakas:1989xu,Pope:1991zka,Pope:1991ig}. This is a classical algebra because we have neglected all the terms involving more than one contraction. If $k=0$ then of course this reduces to an infinite dimensional Abelian algebra. In the context of tree level soft graviton theorems $\wedge w_{1+\infty}$ has already appeared as a symmetry of the dual Celestial CFT$_2$ \cite{Strominger:2021mtt,Adamo:2021lrv}.

\subsection{Conformal transformation of higher spin currents}\label{confhs}
We start with the current $\bar d^{j}$ defined in \eqref{eq:d_j_def}, which appears in the singular part of the Ward identity for the $\mathcal{O}\left(\omega^{2j-1}(\ln\omega)^{2j}\right)$ soft theorem in \eqref{eq:singular_Ward_identity}. Contracting this current with $2j$ insertions of $\bar W$, we obtain
\ben\label{eq:djm_series}
\bar W^{\alpha_1}\bar W^{\alpha_2}\cdots \bar W^{\alpha_{2j}}\, \bar d^j_{\alpha_1\alpha_2\cdots\alpha_{2j}}(\bar z)
&=&\sum_{m=-j}^j {}^{2j}C_{j+m}\bar w^{j+m}\, :\left[\left(\bar d^\f{1}{2}_{\f{1}{2}}(\bar z)\right)^{j+m}\left(\bar d^\f{1}{2}_{-\f{1}{2}}(\bar z)\right)^{j-m} \right]:\non\\
&\equiv &\sum_{m=-j}^j {}^{2j}C_{j+m}\quad \bar d^j_m(\bar z)\, \bar{w}^{j+m},
\een
where the last line defines $\bar d^j_m(\bar z)$ with  $m=\sum\limits_{i=1}^{2j}\alpha_i$ in relation to the LHS, and it's explicit expression is given by
\be
\bar d^j_m(\bar z)\equiv\quad :\left[\left(\bar d^\f{1}{2}_{\f{1}{2}}(\bar z)\right)^{j+m}\left(\bar d^\f{1}{2}_{-\f{1}{2}}(\bar z)\right)^{j-m} \right]:\, ,
\ee
for $-j\leq m\leq j$. In relation to the dipole operator $\bar d(\bar w,\bar z)$ with correlation function \eqref{eq:dipole_op_corr}, the RHS of \eqref{eq:djm_series} can also be identified as normal ordering of $2j$ number of dipole operator, i.e. 
\be
\sum_{m=-j}^j {}^{2j}C_{j+m}\quad \bar d^j_m(\bar z)\, \bar{w}^{j+m} =:\left[\bar d(\bar w,\bar z)\right]^{2j}:\, .
\ee

Now under the $SL(2,\mathbb{R})_{\text{R}}$ transformation described in section \ref{S:algebra}, we find the following transformation rule
\be
:\left[\bar d^\prime(\bar w,\bar z)\right]^{2j}:=\f{(c\bar w+d)^{2j}}{(c\bar z+d)^{2j}} :\left[\bar d(\bar w',\bar z')\right]^{2j}:\, .
\ee 
This in turn provides the following relation for transformation of $\bar d^j_m$:
\ben
\sum_{m=-j}^j {}^{2j}C_{j+m}\bar w^{j+m} \bar d^{\prime j}_m(\bar z)=\f{(c\bar w+d)^{2j}}{(c\bar z+d)^{2j}}  \sum_{n=-j}^j {}^{2j}C_{j+n} \left(\f{a\bar w+b}{c\bar w+d}\right)^{j+n} \bar d^j_n\left(\f{a\bar z+b}{c\bar z+d}\right).\label{eq:djm_recursion}
\een
Now focusing on infinitesimal $SL(2,\mathbb{R})_{\text{R}}$ transformations as in section \ref{S:algebra} we find the following commutation relations
\ben\label{eq:L_djm_comm}
\begin{gathered}
\left[\bar L_{-1},\bar d^j_m(\bar z)\right]=\bar\p \bar d^j_m(\bar z)+(j-m)\bar d^j_{m+1},\\
\left[\bar L_{0},\bar d^j_m(\bar z)\right]=\left(j+m+\bar z\bar\p\right)\bar d^j_m(\bar z),\\
\left[\bar L_{1},\bar d^j_m(\bar z)\right]=\left(2j\bar z+\bar z^2\bar\p\right)\bar d^j_m(\bar z) -(j+m)\bar d^j_{m-1}(\bar z),
\end{gathered}
\een
for $j\in \f{1}{2}\mathbb{Z}_+$ and $-j\leq m\leq j$. This coincides with the algebra given in \eqref{eq:SL2R_commutation} for $j=\f{1}{2}$.

\section{What is the value of $k$?}\label{kk}

The reader must have realized the importance of the value of $k$. If $k$ is zero then the algebra of the higher spin current becomes Abelian. However, if $k$ is nonzero but small such that we can neglect the quantum corrections then the algebra of the higher spin currents becomes the well known $\wedge w_{1+\infty}$. Now in order to know the value of $k$ we need to calculate celestial correlators or $S$-matrix elements of the form 
\be\label{k}
\left\langle\bar d^{j_1}(\bar w_1) \bar d^{j_2}(\bar w_2) \prod_{i=1}^N \phi_{\D_i}^{(\eta_i)}(z_i,\bar z_i,e_i)\right\rangle.
\ee
However, we do not know the Feynman diagram representation of the $S$-matrix elements of the form \eqref{k} because \textit{the higher spin currents do not appear as asymptotic states in a scattering experiment in an obvious way}. 

Now one may think that the (consecutive) double soft theorem is a way out. But, the problem is that the currents appear in the soft theorems only as composite operators of the form $:\bar d^j(\bar z) \phi_{\D}^{(\eta)}(z,\bar z,e):$. If we want to derive the consecutive double soft theorem using the Celestial CFT then we encounter correlation functions with multiple insertions of the composite operators, for example, 
\be\label{comp}
\left\langle:\bar d^{j_1}(\bar z_1)\phi_{\D_1}^{(\eta_1)}(z_1,\bar z_1,e_1): \ :\bar d^{j_2}(\bar z_2)\phi_{\D_2}^{(\eta_2)}(z_2,\bar z_2,e_2): \ \prod_{i=3}^N \phi_{\D_i}^{(\eta_i)}(z_i,\bar z_i,e_i)\right\rangle.
\ee
This correlation function depends on $k$ but to compute it we also need the celestial OPE between the matter primaries $\phi_{\D_1}^{(\eta_1)}(z_1,\bar z_1,e_1)$ and $\phi_{\D_2}^{(\eta_2)}(z_2,\bar z_2,e_2)$. This of course depends on the theory under consideration. Therefore the value of $k$ may differ from theory to theory or it may be zero for every theory. We leave a more decisive statement on the value of $k$ to future research.

\section*{Acknowledgment} We are grateful to Andrew Strominger for pointing out a conceptual problem in the previous version of this paper. We are also grateful to Alok Laddha, Andrea Puhm, Ana-Maria Raclariu and Ashoke Sen for very helpful discussions. SB would also like to thank all the participants of the Pre-ISM Workshop in TIFR Mumbai, where this work was presented, for interesting questions and comments. We acknowledge the use of ChatGPT for providing general information and for assisting with exploratory checks to arrive at the projection operator \eqref{projection}. The work of SB is supported by the  Swarnajayanti Fellowship (File No- SB/SJF/2021-22/14) of the Department of Science and Technology and ANRF, India. The work of RM is supported by ANRF-CRG project (CRG/2022/006165).
The work of BS was supported by STFC under grant ST/X000753/1. 

\appendix
\section{Support for the correctness of conjectured $n$-loop soft factor}\label{app:n-loop_soft_photon}\label{S:appendix}
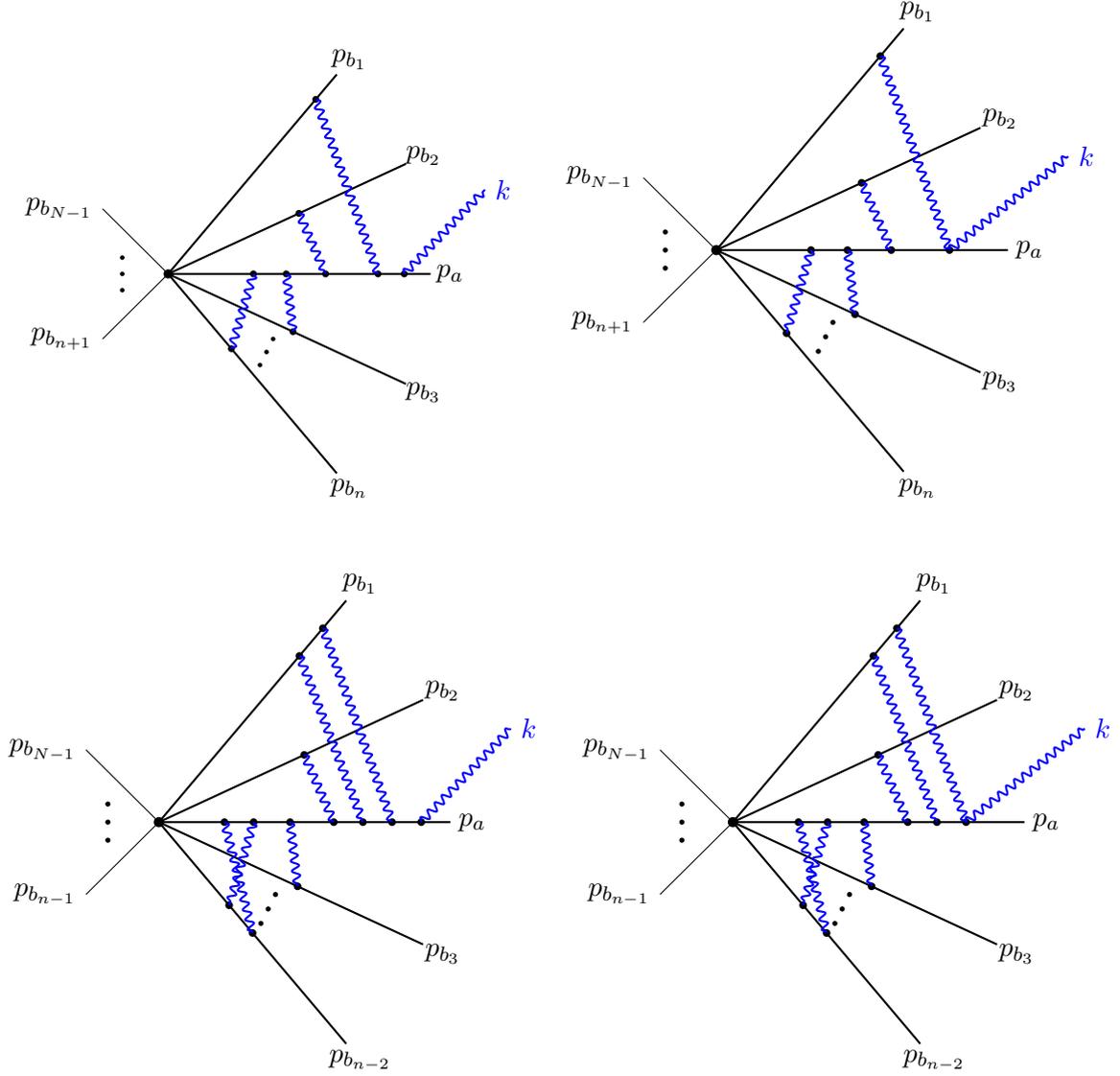
\begin{figure}[h!]
\begin{center}
\begin{tikzpicture}[scale=0.9,
    every node/.style={font=\small},
    photon/.style={decorate, decoration={snake,segment length=4,amplitude=1.5},blue}
]
\coordinate (O) at (0,0);
\fill (O) circle (2pt);
\coordinate (O1) at (-0.7,0.25);
\fill (O1) circle (1pt);
\coordinate (O2) at (-0.7,0);
\fill (O2) circle (1pt);
\coordinate (O3) at (-0.7,-0.25);
\fill (O3) circle (1pt); 

\coordinate (O4) at (1.6,-1);
\fill (O4) circle (1pt);
\coordinate (O5) at (1.5,-1.2);
\fill (O5) circle (1pt);
\coordinate (O6) at (1.4,-1.4);
\fill (O6) circle (1pt);

\draw (O)--(-1,-1) node[left] {$p_{b_{n+1}}$};
\draw (O)--(-1,1) node[left] {$p_{b_{N-1}}$};

\foreach \i/\ang/\lab in {
    1/50/{p_{b_1}},
    2/25/{p_{b_2}},
    3/0/{p_{a}},
    4/-25/{p_{b_3}},
    5/-50/{p_{b_n}}
}{

    \draw[thick] (O) -- ++(\ang:4);

    \ifx\lab\empty
    \else
        \node at (\ang:4.3) {$\lab$};
    \fi
}

\fill (50:3.5) circle (1.5pt);
\fill (0:3.2) circle (1.5pt);
\draw[photon,thick]
 (50:3.5)  -- (0:3.2);

\fill (25:2.2) circle (1.5pt);
\fill (0:2.4) circle (1.5pt);
\draw[photon,thick]
    (25:2.2) -- (0:2.4);

\fill (0:1.8) circle (1.5pt);
\fill (-25:2.1) circle (1.5pt);
\draw[photon,thick]
    (0:1.8) -- (-25:2.1);
    
    \fill (0:1.3) circle (1.5pt);
\fill (-50:1.5) circle (1.5pt);
\draw[photon,thick]
    (0:1.3) -- (-50:1.5);

\fill (0:3.6) circle (1.5pt);
 (0:3.6)
\draw[photon,thick]
 (0:3.6)--(15:5.0) node[right]{$k$};
\end{tikzpicture}
\quad
\begin{tikzpicture}[scale=1.0,
    every node/.style={font=\small},
    photon/.style={decorate, decoration={snake,segment length=4,amplitude=1.5},blue}
]

\coordinate (O) at (0,0);
\fill (O) circle (2pt);
\coordinate (O1) at (-0.7,0.25);
\fill (O1) circle (1pt);
\coordinate (O2) at (-0.7,0);
\fill (O2) circle (1pt);
\coordinate (O3) at (-0.7,-0.25);
\fill (O3) circle (1pt); 

\coordinate (O4) at (1.6,-1);
\fill (O4) circle (1pt);
\coordinate (O5) at (1.5,-1.2);
\fill (O5) circle (1pt);
\coordinate (O6) at (1.4,-1.4);
\fill (O6) circle (1pt);

\draw (O)--(-1,-1) node[left] {$p_{b_{n+1}}$};
\draw (O)--(-1,1) node[left] {$p_{b_{N-1}}$};

\foreach \i/\ang/\lab in {
    1/50/{p_{b_1}},
    2/25/{p_{b_2}},
    3/0/{p_{a}},
    4/-25/{p_{b_3}},
    5/-50/{p_{b_n}}
}{

    \draw[thick] (O) -- ++(\ang:4);

    \ifx\lab\empty
    \else
        \node at (\ang:4.3) {$\lab$};
    \fi
}

\fill (50:3.5) circle (1.5pt);
\fill (0:3.2) circle (1.5pt);
\draw[photon,thick]
 (50:3.5)  -- (0:3.2);

\fill (25:2.2) circle (1.5pt);
\fill (0:2.4) circle (1.5pt);
\draw[photon,thick]
    (25:2.2) -- (0:2.4);

\fill (0:1.8) circle (1.5pt);
\fill (-25:2.1) circle (1.5pt);
\draw[photon,thick]
    (0:1.8) -- (-25:2.1);
    
    \fill (0:1.3) circle (1.5pt);
\fill (-50:1.5) circle (1.5pt);
\draw[photon,thick]
    (0:1.3) -- (-50:1.5);

\draw[photon,thick]
 (0:3.2)--(15:5.0) node[right]{$k$};
\end{tikzpicture} \linebreak

\begin{tikzpicture}[scale=1.0,
    every node/.style={font=\small},
    photon/.style={decorate, decoration={snake,segment length=4,amplitude=1.5},blue}
]

\coordinate (O) at (0,0);
\fill (O) circle (2pt);
\coordinate (O1) at (-0.7,0.25);
\fill (O1) circle (1pt);
\coordinate (O2) at (-0.7,0);
\fill (O2) circle (1pt);
\coordinate (O3) at (-0.7,-0.25);
\fill (O3) circle (1pt); 

\coordinate (O4) at (1.6,-1);
\fill (O4) circle (1pt);
\coordinate (O5) at (1.5,-1.2);
\fill (O5) circle (1pt);
\coordinate (O6) at (1.4,-1.4);
\fill (O6) circle (1pt);

\draw (O)--(-1,-1) node[left] {$p_{b_{n-1}}$};
\draw (O)--(-1,1) node[left] {$p_{b_{N-1}}$};

\foreach \i/\ang/\lab in {
    1/50/{p_{b_1}},
    2/25/{p_{b_2}},
    3/0/{p_{a}},
    4/-25/{p_{b_3}},
    5/-50/{p_{b_{n-2}}}
}{

    \draw[thick] (O) -- ++(\ang:4);

    \ifx\lab\empty
    \else
        \node at (\ang:4.3) {$\lab$};
    \fi
}

\fill (50:3.5) circle (1.5pt);
\fill (0:3.2) circle (1.5pt);
\draw[photon,thick]
 (50:3.5)  -- (0:3.2);
 
 \fill (50:3.0) circle (1.5pt);
\fill (0:2.8) circle (1.5pt);
\draw[photon,thick]
 (50:3.0)  -- (0:2.8);

\fill (25:2.2) circle (1.5pt);
\fill (0:2.4) circle (1.5pt);
\draw[photon,thick]
    (25:2.2) -- (0:2.4);

\fill (0:1.8) circle (1.5pt);
\fill (-25:2.1) circle (1.5pt);
\draw[photon,thick]
    (0:1.8) -- (-25:2.1);
    
    \fill (0:1.3) circle (1.5pt);
\fill (-50:1.5) circle (1.5pt);
\draw[photon,thick]
    (0:1.3) -- (-50:1.5);
    
      \fill (0:0.9) circle (1.5pt);
\fill (-50:2.0) circle (1.5pt);
\draw[photon,thick]
    (0:0.9) -- (-50:2.0);

\fill (0:3.6) circle (1.5pt);
 (0:3.6)
\draw[photon,thick]
 (0:3.6)--(15:5.0) node[right]{$k$};
\end{tikzpicture}\quad
\begin{tikzpicture}[scale=1.0,
    every node/.style={font=\small},
    photon/.style={decorate, decoration={snake,segment length=4,amplitude=1.5},blue}
]

\coordinate (O) at (0,0);
\fill (O) circle (2pt);
\coordinate (O1) at (-0.7,0.25);
\fill (O1) circle (1pt);
\coordinate (O2) at (-0.7,0);
\fill (O2) circle (1pt);
\coordinate (O3) at (-0.7,-0.25);
\fill (O3) circle (1pt); 

\coordinate (O4) at (1.6,-1);
\fill (O4) circle (1pt);
\coordinate (O5) at (1.5,-1.2);
\fill (O5) circle (1pt);
\coordinate (O6) at (1.4,-1.4);
\fill (O6) circle (1pt);

\draw (O)--(-1,-1) node[left] {$p_{b_{n-1}}$};
\draw (O)--(-1,1) node[left] {$p_{b_{N-1}}$};

\foreach \i/\ang/\lab in {
    1/50/{p_{b_1}},
    2/25/{p_{b_2}},
    3/0/{p_{a}},
    4/-25/{p_{b_3}},
    5/-50/{p_{b_{n-2}}}
}{

    \draw[thick] (O) -- ++(\ang:4);

    \ifx\lab\empty
    \else
        \node at (\ang:4.3) {$\lab$};
    \fi
}

\fill (50:3.5) circle (1.5pt);
\fill (0:3.2) circle (1.5pt);
\draw[photon,thick]
 (50:3.5)  -- (0:3.2);
 
 \fill (50:3.0) circle (1.5pt);
\fill (0:2.8) circle (1.5pt);
\draw[photon,thick]
 (50:3.0)  -- (0:2.8);

\fill (25:2.2) circle (1.5pt);
\fill (0:2.4) circle (1.5pt);
\draw[photon,thick]
    (25:2.2) -- (0:2.4);

\fill (0:1.8) circle (1.5pt);
\fill (-25:2.1) circle (1.5pt);
\draw[photon,thick]
    (0:1.8) -- (-25:2.1);
    
    \fill (0:1.3) circle (1.5pt);
\fill (-50:1.5) circle (1.5pt);
\draw[photon,thick]
    (0:1.3) -- (-50:1.5);
    
      \fill (0:0.9) circle (1.5pt);
\fill (-50:2.0) circle (1.5pt);
\draw[photon,thick]
    (0:0.9) -- (-50:2.0);

\draw[photon,thick]
 (0:3.2)--(15:5.0) node[right]{$k$};
\end{tikzpicture}
\captionsetup{font=small}
\caption{$n$-loop Feynman diagrams contributing to $\mathbb{S}_{\text{em}}^{(n-1)}$ in the theory of massless scalar QED. }\label{f:Feynman}
\end{center}
\end{figure}

Building on the derivation of the $\mathcal{O}(\omega(\ln\omega)^2)$ soft photon theorem in
four-dimensional massive scalar QED with scalar contact interactions,
obtained from an explicit analysis of two-loop Feynman diagrams in
\cite{Sahoo:2020ryf}, we revisit the result at the leading order in the high-energy limit given in \eqref{domain}. We find that, after performing the loop integrals and
subsequently taking the high-energy limit \eqref{domain}, the only
non-vanishing contributions at the leading order arise from diagrams in which one end of each of
the two virtual photon propagators is attached to the same charged particle
line that emits the soft photon. All remaining two-loop diagrams, some of which do
contribute to the $\mathcal{O}(\omega(\ln\omega)^2)$ soft factor at general kinematics of charged
particle scattering does not contribute to the leading order in the high-energy expansion \eqref{domain}. After examining a few higher-loop orders, this feature appears to be generic for leading order contribution to soft photon factor in the high-energy limit \eqref{domain}. Now assuming this observation to be valid for all-loop order S-matrix  in scalar QED, we find that the only $n$-loop Feynman diagrams in Fig.\ref{f:Feynman}  provide non-vanishing contributions to $\mathbb{S}_{\text{em}}^{(n-1)}$ at leading order in the high-energy limit \eqref{domain}, when evaluated with G-photon virtual propagators \cite{Sahoo:2018lxl,Sahoo:2020ryf}. The first diagram in Fig.\ref{f:Feynman} represents the set of diagrams in which all the $n$ virtual photon (blue wavy lines) propagators are attached between particle-$a$ and $n$ different charged particles (black solid lines) through scalar–scalar–photon vertices, and the soft photon is attached to particle-$a$ at the far right through a scalar–scalar–photon vertex. The second diagram in Fig.\ref{f:Feynman} represents the set of diagrams in which all but one of the $n$ virtual photon propagators are attached between particle-$a$ and $n$ different charged particles through scalar–scalar–photon vertices, while the soft photon is attached to particle-$a$ at the far right through a scalar–scalar–photon–photon vertex where one of the remaining virtual photon is also attached. The third diagram in Fig.\ref{f:Feynman} represents the set of diagrams that differ from the first diagram only in that two or more virtual photons are now attached between particle-$a$ and particle-$b_i$ via ladder-type or cross-ladder-type configurations. Similarly, the fourth diagram in Fig.\ref{f:Feynman} represents the set of diagrams that differ from the second diagram only in that two or more virtual photons are attached between particle-$a$ and particle-$b_i$ via ladder-type or cross-ladder-type configurations.

From the ratio of two amplitudes as follows from \eqref{eq:Soft_expansion}, the leading-log $n$-loop soft factor at leading order in the high-energy limit \eqref{domain} from the set of the Feynman diagrams in Fig.\ref{f:Feynman} turns out to be
\ben
S^{n-\text{loop}}_{\text{em}}\left(\lbrace e_a,p_a\rbrace ,(\varepsilon^h,k)\right)&=&\f{1}{n!}\sum_{a=1}^N e_a\f{\varepsilon^h_\mu k_\nu}{p_a.k} \left(p_a^\mu \f{\p K_{\text{em}}}{\p p_{a\nu}}-p_a^\nu \f{\p K_{\text{em}}}{\p p_{a\mu}}\right)\left(k_\sigma \f{\p K_{\text{em}}}{\p p_{a\sigma}}\right)^{n-1},\label{eq:Snloop}
\een
where the contributions of first term inside the parenthesis comes from the sum over first and third Feynman diagrams in Fig.\ref{f:Feynman} and  the contributions of second term inside the parenthesis comes from the sum over third and fourth Feynman diagrams in Fig.\ref{f:Feynman}. The expression for $K_{\text{em}}$ is given by
\ben
K_{\text{em}}&=&\f{i}{2}\sum_{\substack{b,c=1\\ c\neq b}}^N e_be_c\, (p_b.p_c) \int_{\omega}^\Lambda \f{d^4\ell}{(2\pi)^4}\f{1}{(\ell^2-i\epsilon)(p_b.\ell+i\epsilon)(p_c.\ell -i\epsilon)}\non\\
&\simeq & \f{i}{2}(\ln\omega)\sum_{\substack{b,c=1\\ c\neq b}}^N \f{e_be_c}{4\pi} \left[\delta_{\eta_b\eta_c,1}+\f{i}{2\pi}\log\left((p_b.p_c)^2\right)\right],
\een
where $\Lambda$ is of order $\sqrt{|p_b.p_c|}$, and in the second line after using the sign $\simeq$ we only kept the $\mathcal{O}(\ln\omega)$ contribution. Following the terminology of \cite{Sahoo:2018lxl}, we refer to the first term inside the parentheses, namely $\delta_{\eta_b\eta_c,1}$, as the ``classical'' contribution, and the second term as the ``quantum'' contribution.\footnote{For a careful analysis of the ``classical'' part of the above integral in the massless case, the reader is referred to the appendix of \cite{Choi:2025mzg}.} Now substituting the expression for $K_{\text{em}}$ in \eqref{eq:Snloop} we find the following soft factor coefficient at order $\omega^{n-1}(\ln\omega)^n$ and at the leading order in the high-energy limit \eqref{domain}:
\ben
\mathbb{S}_{\text{em}}^{(n-1)}= - \frac{1}{n!(4\pi^2)^{n}}\sum_{\substack{a,b=1\\ b\neq a}}^N  \frac{e^2_{a}e_{b}}{p_{a}.p_{b}}\frac{\varepsilon^h_\mu \mathbf{n}_\nu}{p_{a}.\mathbf{n}}\left(p_{a}^\mu p_{b}^\nu -p_{a}^\nu p_{b}^\mu\right)\left( \sum_{\substack{c=1\\ c\neq a}}^N \frac{e_{a}e_{c}}{p_{a}.p_{c}}(p_{c}.\mathbf{n})\right)^{n-1},
\een
which is quoted in \eqref{eq:soft_photon_coeff}.



\end{document}